\documentclass[12pt,preprint]{aastex}
\usepackage{graphicx}

\shorttitle{Two-Pole Caustic Model}
\shortauthors{Dyks and Rudak}

\begin{document}

\title{Two-Pole Caustic Model for High-Energy Lightcurves of Pulsars}

\author{J. Dyks\altaffilmark{1}}
\affil{Laboratory for High Energy Astrophysics, NASA/GSFC, 
        Greenbelt, MD 20771}
        
\email{jinx@milkyway.gsfc.nasa.gov}    
    
\and

\author{B. Rudak}
\affil{Nicolaus Copernicus Astronomical Center, 
        Rabia{\' n}ska 8, 87-100 Toru{\' n}, Poland}
        
\email{bronek@camk.edu.pl}        

\altaffiltext{1}{On leave from Nicolaus Copernicus Astronomical Center, Poland}

\begin{abstract}
We present a new model of high-energy lightcurves from rotation powered pulsars. 
The key ingredient of the model 
is the gap region (i.e. the region where particle
acceleration is taking place and high-energy photons originate) which
satisfies the following assumptions:
i) the gap region extends from each polar cap to the light cylinder;
ii) the gap is thin and confined to the surface of last open 
magnetic-field lines;
iii) photon emissivity is uniform within the gap region.
The model lightcurves are dominated by strong peaks (either double or single) of caustic origin.
Unlike in other pulsar models with caustic effects, 
the double peaks arise due to 
crossing two caustics, each of which
is associated with a different magnetic pole.
The generic features of the lightcurves 
are consistent
with the observed characteristics of pulsar lightcurves: 
1) the most natural (in terms of probability) shape consists of two peaks (separated by 0.4 to 0.5 in phase
for large viewing angles);
2) the peaks posess well developed wings;
3) there is a bridge (inter-peak) emission component;
4) there is a non-vanishing off-pulse emission level;
5) the radio pulse occurs before the leading high-energy peak.
The model is well suited for four gamma-ray pulsars - Crab, Vela, Geminga and B1951+32 - with double-peak lightcurves
exhibiting the peak separation of 0.4 to 0.5 in phase.
Hereby, we apply the model to the Vela pulsar. 
Moreover, we indicate
the limitation of the model in accurate reproducing of the lightcurves with single pulses
and narrowly separated (about 0.2 in phase) pulse peaks. 
We also discuss the optical polarization properties for the Crab pulsar 
in the context of the two-pole caustic model.   
\end{abstract}

\keywords{pulsars: high-energy radiation, lightcurves}

\section{Introduction} 
The striking feature of the lightcurves\footnote{We use the term
`lightcurves' for averaged pulse profiles.}
of known gamma-ray pulsars
are relatively long duty cycles as well as phase shifts in comparison to the radio pulses 
(Thompson et al. 1999, Thompson 2001, Kanbach 2002).
The lightcurve shapes fall into three categories. 
The three brightest gamma-ray pulsars - the Crab pulsar, Vela and Geminga,
along with B1951+32
exhibit two well defined, sharp peaks separated in phase by $0.4 - 0.5$, and connected
by the interpeak bridge of a considerable level.
B1706-44 and B1055-52 show two peaks separated by about 0.2 in phase, whereas B1509-58 exhibits a broad single pulse. 
Similar properties are present in the X-ray domain of the high-energy
emission.
Moreover, in the case of the Crab pulsar its optical pulsed emission is
usually
considered jointly with the high-energy emission:
the phase-averaged spectra in gamma and X-rays connect smoothly to
optical, and the pulses have similar shapes and phases.
This suggests that gammas, X-rays and optical light may come from the
same regions.

These properties, along with the spectral properties which are not 
the subject of this paper, prompted substantial refinements of
 the physical models
of pulsar high-energy activity, both outer-gap models (Cheng et al.~1986,
Romani \& Yadigaroglu 1995, Zhang \& Cheng 1997) 
as well as polar-cap models (Sturrock 1971; Ruderman \& Sutherland
1975; Daugherty \& Harding 1982, Sturner et al.~1995),
(see Rudak, Dyks \& Bulik 2002 for recent critical review).
Despite reiterated arguments in favour of polar-gap models 
(e.g. Baring 2001)
and outer-gap models (e.g. Yadigaroglu \& Romani 1995),
both classes of models
still suffer from serious problems.
According to the polar cap model, the characteristic double-peak 
lightcurve forms when the line of sight intersects the
polar-cap beam where the highest-energy emission originates: upon entering the beam
a leading peak is produced, followed by 
a bridge emission due to the inner parts of the beam, when the line of sight exits the beam
the trailing peak forms.
Because of long duty cycles in the high-energy domain
on one hand, and narrow opening angles for gamma-ray emission on the other hand, polar cap
models have to rely on nearly aligned rotators,
where inclination (the angle $\alpha$) of the magnetic axis (the magnetic moment $\vec \mu$)
to the spin axis (the angular velocity $\vec \Omega$) is  
comparable to the angular extent of the polar cap (Daugherty \& Harding
1994). 

The outer-gap model, on the contrary, prefers highly inclined rotators 
(Chiang \& Romani 1992; Romani \& Yadigaroglu 1995; Cheng et al. 2000).
However, the model 
is unable (without additional postulates) to account 
for the presence of outer wings in the 
double-peak lightcurves. More importantly, the model in its present form 
(e.g. Zhang \& Cheng 2002) is unable to
account for a substantial level of the off-pulse emission in the Crab pulsar.
Also on theoretical grounds the existence of outer gaps in the present-day `vacuum' approach (i.e. with
the gap extending between the null surface and the light cylinder) has been questioned.
The outer gap remains anchored to the conventional null surface, provided that
no current is injected at the boundaries of the accelerator (no pairs are created); otherwise
the gap extention becomes very sensitive to the details of pair creation
(Hirotani \& Shibata 2001, Hirotani et al. 2003).

These problems motivated us to propose a new picture of the origin of high-energy radiation within the
pulsar magnetospheres - in regions confined to the surface of last open magnetic field lines
(similarly to thin outer-gap accelerators) but extending between the polar cap and the light cylinder.
The most important consequence of such extended accelerators, as far as the lightcurves are concerned,
is a caustic nature of the high-energy peaks: 
special relativity effects
(aberration of photon emission directions and time of flight delays
due to the finite speed of light $c$) cause that photons emitted at
different altitudes within some regions of the magnetosphere
are piled up at the same phase of a pulse (Morini 1983; Romani \&
Yadigaroglu 1995).

The term `high-energy radiation' used throughout the paper refers to nonthermal
radiation in the energy domain of gamma-rays
and hard X-rays (i.e. above several keV). The soft X-ray radiation ($0.1 \la E \la 1$ keV) 
is not included into our considerations because 
it is heavily affected by thermal emission from the neutron star surface in some objects, like Geminga 
which exhibits a complex pattern of soft X-ray pulses 
(e.g. fig.4 in Jackson et al.~2002).
For reasons mentioned at the beginning of this section, in the case of the Crab pulsar our definition 
includes also the optical band.

The paper is arranged in the following way: In section 2 we
introduce the two-pole caustic model for the high-energy lightcurves of pulsars and present the results of numerical calculations.
Section~3 contains discussion of the generic features of the model as well as comparison with the properties of other models.

\section{The two-pole caustic model}        
Special relativity (SR) effects which affect pulsar lightcurves
include the aberration of photon emission directions and the time of flight
delays caused by the finite speed of light $c$.
Morini (1983) was the first to prove that these effects
were
able to produce prominent peaks in pulsar lightcurves.
He obtained the peaks of caustic
origin
in his version of the polar cap model, which included
photon emission from high altitudes, where
the SR effects are important.
Unlike in the case of the Morini's model,
in the standard polar cap model the strongest 
gamma-ray emission takes place
very close to the star surface, where the caustic effects are not
important and, therefore, pulsar 
lightcurves are mostly determined
by the altitudinal extent of the accelerator (polar gap) 
and by the emissivity
profile along the magnetic field lines (eg.~Daugherty \& Harding 1996;
Dyks \& Rudak 2000).

In the recent version of the outer gap model (Chiang \& Romani 1992;
Romani \& Yadigaroglu 1995; Yadigaroglu 1997; Cheng et al.~2000) the peaks in 
the lightcurves are purely due to the caustic effects -- the fact first
emphasized by Romani \& Yadigaroglu (1995).
However,
the altitudinal extent of accelerator (limited by the position of the inner boundary
of the outer gap)  is crucial 
in limiting the possible 
shapes of lightcurves within this model.  
Interestingly, Yadigaroglu (1997) also considered 
photon emission along the entire length of all last open magnetic field
lines, relaxing thus the outer-gap extent limit  (see the bottom panel in Fig.~3.1 of his thesis). 
However, he didn't pursue this possibility in any further details.

Hereby we propose that the observed high-energy emission from
known gamma-ray pulsars originates from the regions considered already by 
Yadigaroglu (1997).
Let us assume that the gap region (the region where particle acceleration is taking place as well as
high-energy photons originate) posesses the following properties: \\
- the gap region extends from the polar cap to the light cylinder,\\
- the gap is thin and confined to the surface of last open magnetic-field lines,\\
- photon emissivity is uniform within the gap region.\\
Fig.1 shows schematically the location and the extension of the proposed gap (for the sake of comparison
the location of the outer gap is shown as well).
The resulting  lightcurves are dominated by strong peaks (either double or single) of caustic origin.

In the outer-gap model, the inner boundary of the gap is 
located at an intersection of the null-charge surface with the last closed field lines.
Its radial distance $r_{\rm in}$ is, therefore, a function of azimuthal angle $\varphi$, with a minimum
value $r_{\rm in, min}$ in the $(\vec \Omega, \vec \mu)$-plane.
For highly inclined rotators $r_{\rm in, min}$ becomes a small fraction
of the light cylinder radius $r_{\rm lc} = c/\Omega$: $r_{\rm in, min}/r_{\rm lc} \approx [2/(3\tan \alpha)]^2$ 
(Halpern \& Ruderman 1993);
for the inclination angle $\alpha = 60^\circ$ the ratio is $0.15$.  However, for azimuthal directions departing
the $(\vec \Omega, \vec \mu)$-plane, the inner radius $r_{\rm in}$ tends to $r_{\rm lc}$.
The radiation region extends out
to the light cylinder and the radiation escaping the magnetospheric region comes from particles 
moving outwards along the magnetic lines.\\ 
In our model we assume, however,
that the actual gap extends to the polar cap, i.e. $r_{\rm in} \sim r_{\rm ns}$ for all azimuthal angles $\varphi$.
This assumption is essential for the expected performance of the model since it implies that the observer can detect radiation
originating from \emph{both} magnetic hemispheres.

As in the outer-gap model of Chiang \& Romani (1992), and Cheng et al. (2000) we assumed that the photon emissivity is uniform 
everywhere in the proposed gap region.
For simplicity, we consider a rigidly rotating static-like magnetic dipole.  
Departures of the retarded field lines from
the static case are of the order of $\beta^2$ 
and they are insignificant since prominent features in modelled
lightcurves (peaks) which we discuss below
arise due to radiation at radial distances 
$r < 0.75 \, r_{\rm lc}$.
Also, rotationally-driven currents can be neglected:
longitudinal currents suspected to flow within the open field line region
cannot modify $\vec B$ by a factor exceeding $\beta^{3/2}$ whereas
toroidal currents due to plasma corotation
change $\vec B$ barely by $\beta^2$ 
($\vec\beta$ is the local corotation velocity in the speed 
of light units; $\beta=|\vec\beta|$).

We considered radiating particles moving from the outer rim of the polar cap
toward the light cylinder.
The photons were emitted tangentially to local magnetic field lines 
in the corotation frame, and then
they were followed crossing the magnetosphere with no magnetic attenuation.
Our treatment of rotational effects in the numerical code 
used for the calculations is the same as in Yadigaroglu (1997)
and in Dyks \& Rudak (2002). 
Specifically, we
used the following standard aberration formula to transform the unit vector of photon propagation 
direction $\hat \eta^\prime$ from the corotating frame 
to its value $\hat \eta$ in the inertial observer frame:
\begin{equation}
\hat\eta = \frac{\hat\eta^\prime + (\gamma +
(\gamma-1)(\vec\beta\cdot\hat\eta^\prime)/\beta^2)\ \vec\beta}
{\gamma(1 + \vec\beta\cdot\hat\eta^\prime)},
\end{equation}
where $\gamma = (1 - \beta^2)^{-1/2}$.
The above formula results directly from the general Lorentz
transformation. By replacing $\hat\eta$ and $\hat\eta^\prime$
in the formula with $\vec v c^{-1}$ and $\vec v^\prime c^{-1}$ (respectively),
one obtains a general Lorentz transformation for a velocity vector $\vec v$.\\
Photon travel
delays were taken into account by adding 
$\vec r \cdot \hat \eta/r_{\rm lc}$
to the azimuth $\phi_{\rm em}$ 
of photon emission direction $\hat \eta$
($\vec r$ is the radial position of emission point, and $\hat \eta$ is the unit vector
of photon propagation direction after aberration effect is
included).
The result, taken with a minus sign, is a phase of detection $\phi$:
\begin{equation}
\phi = -\phi_{\rm em} - \vec r \cdot \hat \eta/r_{\rm lc}.
\end{equation}
Our numerical code performs a raytracing followed by a numerical
integration of the observed pulse profile.
All results described below are just due to
the dipolar shape of the magnetic field, the aberration effect
(eq.~1), as well as the light travel time delays (eq.~2).

The results are presented in Fig.\ref{fig2} in the form of 
photon mapping onto the $(\zeta_{\rm obs}, \phi)$-plane, with accompanying lightcurves for five
viewing angles $\zeta_{\rm obs}$ (the angle between the spin axis and the line of sight).
Inclination angle $\alpha = 60^\circ$, and spin period $P = 0.033$~s were assumed for the rotator.
For each magnetic pole two caustics form in the $(\zeta_{\rm obs}, \phi)$-plane:
1) the dominant caustic, and 2) the subdominant caustic.\\
The dominant caustic (easy to identify in the photon mapping) is associated with
the trailing part of the emission region with respect to its magnetic pole.
The subdominant caustic (much weaker) is associated with the leading part 
of the emission region.
Characteristic features in the lightcurves due to caustic crossing 
for large values of $\zeta_{\rm obs}$ are marked with 
capital letters from A to D. 
The dominant-caustics crossing yields two prominent peaks D and B, 
in the lightcurve.
The peaks consist of photons emitted over a very wide range of altitudes
(e.g.~for $\alpha \sim \zeta_{\rm obs} \sim 90^\circ$ almost all altitudes
between the star surface and the light cylinder contribute to the peaks;
see also fig.~2
in Morini 1983).
The subdominant-caustics crossing yields the features A and C, 
which actually contribute
to the trailing wings of the peaks D and B, respectively.
Features A and C consist of photons emitted very close to the
light cylinder (cf.~fig.~9 in Cheng et al.~2000).
This result is in clear contrast with the results obtained 
for the conventional outer-gap model (see Fig.\ref{fig3}), where
the leading peak (A) is due to the subdominant caustic, and 
the trailing peak (B) is due to the dominant caustic. 
Unlike in the outer gap model, in our model each of the two prominent peaks
in the resulting lightcurve is associated with a different magnetic pole.
Therefore, we propose the name ``two-pole caustic model".

The position of the last closed magnetic field lines
and the magnitude of special relativity effects 
are governed by the proximity 
of the light cylinder and, therefore, the radiation pattern as well as
the lightcurves shown in Fig.~2 do not depend on the rotation period $P$ but solely
on the inclination angle $\alpha$
(with one exception: the size of blank
spots corresponding to polar caps does depend on $P$).
Therefore, the two-pole caustic model is relevant for pulsars with any rotation
period. 
We find that noticeable peaks
of caustic origin appear in the lightcurves 
practically for any
inclination of magnetic dipole $\alpha \ne 0$ and for any viewing angle
$\zeta_{\rm obs}\ne 0$. 
For small inclinations ($\alpha \la 30^\circ$) 
the peaks are broader and it is more probable 
to observe single-peaked lightcurves; such lightcurves form for a wide
range of viewing angles for which the coventional outer gap model
predicts no emission. In the GLAST era, the detectability 
of ``moderately inclined pulsars" viewed far from the equatorial plane will serve
as a clear discriminator between the two-pole caustic model and the traditional
outer gap model.

Fig.~\ref{fig4} presents a comparison of a lightcurve 
calculated within the
two-pole caustic model for the Vela pulsar with a lightcurve obtained
for this pulsar with EGRET (Kanbach 1999). 
The model lightcurve was calculated for
electrons distributed evenly along the polar cap rim; 
the density profile across the rim was assumed to be the Gaussian function $f(\theta_{\rm m})$, symmetrical
about $\theta_{\rm m} = \theta_{\rm pc}$, with $\sigma = 0.025\theta_{\rm pc}$
($\theta_{\rm m}$ is
the magnetic colatitude of magnetic field lines' footprints at the star surface,
and $\theta_{\rm pc} \approx \sqrt(r_{\rm ns}/r_{\rm lc})$ is the  magnetic colatitude of the rim).
Photon emission was followed up to $r_{\rm max} = 0.95r_{\rm lc}$.
The shape of the EGRET lightcurve is very well reproduced by the
two-pole caustic model: the leading peak is narrower than the trailing peak,
which connects smoothly with the bridge 
emission;
the leading peak seems to present a separate entity -- it does not
connect smoothly with the bridge, and is followed by a characteristic `interpulse
bump'. 
These features result generically from the two-pole caustic model, since
it predicts that the trailing peak, the bridge emission, and the
`interpulse bump' arise from sampling a single, continuous radiation
pattern from one magnetic pole (cf.~Fig.~\ref{fig2}a). 
The leading peak and the offpulse emission are produced by sampling an
emission pattern from the opposite pole.
The `interpulse bump' 
predicted by the two-pole caustic model is not observed
in the Crab pulsar. This might be caused by a decline in the photon emissivity
above $r \sim 0.5r_{\rm lc}$.

Single peaked gamma-ray lightcurves as the one observed for B1509$-$58 (Kuiper et al.~1999)
can be reproduced in the two-pole caustic model for small viewing angles
(Fig.~\ref{fig2}b), however, the predicted 
phase lag between the gamma-ray and the radio peak ($\sim 0.1$)
is smaller than the one
observed for B1509$-$58 ($\sim 0.35$). 
Moreover, double-peak lightcurves with small separation between
the peaks in B1706$-$44 (Thompson et al.~1996), and B1055$-$52 (Thompson et al.~1999) cannot be interpreted 
in the same way as the lightcurves with wide separation of the peaks.

These difficulties forced Chiang \& Romani (1992) (and probably
Yadigaroglu 1997) to abandon the
geometry of the two-pole caustic model (R.~W.~Romani, private communication).
We agree with R.~W.~Romani
that these particular lightcurves should be interpreted in terms of
the outer gap caustics A and B (Figs.~\ref{fig2} and \ref{fig3}).
This interpretation can be accommodated by the two-pole caustic model only when
the outer gap part of the lightcurves (between A and B in Fig.~\ref{fig2})
dominates over the leading peak formed by the trailing caustic D.
As can be noticed in Fig.~\ref{fig2}, the intensity of the outer gap part
of lightcurves (between A and B) increases with respect to the intensity
of the peak D, as the viewing angle $\zeta_{\rm obs}$
approaches the value for which the line of sight barely
skims the outer gap part of the radiation pattern.
For some viewing gometries the intensity of the outer gap part (A--B) may 
exceed by a few times the intensity of the peak D.
Fig.~\ref{fig5} shows an example of such a lightcurve.
Given the low intensity and large width of the peak D, it may stay
unresolved in the low statistics data of B1509$-$58, B1706$-$44,
and B1055$-$52.

Table~1 summarizes major similarities and differences between our model and other models.
For these purposes we choose the model of Morini 1983 (it was the first model where 
caustic effects were noticed) and the model of Smith et al. 1988, along with
the polar-cap model and the outer-gap model.

\section{Discussion}      
We introduced a two-pole caustic model for the high-energy lightcurves of pulsars
(for the Crab pulsar in particular).
The effects of aberration and light travel delays,
as well as the geometry of the last closed magnetic field lines
are essential for forming the lightcurves of a caustic nature.
The generic features in the lightcurves provided by the two-pole caustic model are consistent
with the observed characteristics
of pulsar lightcurves:\\
1. the most natural (in terms of probability) shape consists of two peaks (separated by 0.4 to 0.5 in phase
for large viewing angles),\\
2. the peaks posess well developed wings,\\
3. there is a bridge (inter-peak) emission component,\\
4. there is a non-vanishing off-pulse emission level,\\
5. the radio pulse (or pulse precursor, in the case of Crab)\footnote{Following the traditional approach (e.g. Yadigaroglu 1997) we
attribute
the single radio pulse (or the precursor in the case of Crab)
to emission associated with the magnetic poles; that determines phase
zero for high-energy lightcurves.}
comes ahead of the leading peak 
(by $\sim 0.1$ in phase for large viewing angles).

Features 1., 3., and 5. are not a unique property of the two-pole caustic model, they can be easily obtained 
within the outer-gap model (compare
the lightcurves in Figs.\ref{fig2} and \ref{fig3}).
Feature 2. may in principle be obtained within the outer-gap model, but no consensus exists
among the proponents of that model on the actual physical reason behind this feature (Yadigaroglu 1997, 
Cheng, Ruderman \& Zhang 2000).
In our model the trailing wings are formed by the subdominant
caustics (A) and (C) which often blend with peaks (D) and (B), respectively 
(the effect of blending is not shown in Fig.~2 since the calculation were truncated
at $0.75r_{\rm lc}$.) 
Feature 4., however, may play a decisive role in 
showing the advantage of the two-pole caustic model over the outer-gap model:
this particular feature of our model may 
explain the presence of the significant X-ray flux from
the Crab pulsar at pulse minimum discovered by Tennant et al.~\cite{t01}.
It has not been demonstrated so far
how such a feature could be obtained within the outer-gap model; in particular,
it is absent in the X-ray pulse profile calculated for the Crab pulsar
in the recent model of Zhang \& Cheng 2002. 

An interesting property of the double-peak lightcurve, 
inherent to the two-pole caustic model,
emerges for viewing angles $\zeta_{\rm obs}$ close to  $90^\circ$: 
the trailing peak (together with its wings) 
assumes the shape which is roughly 
similar (in the sense of translations in the rotation-phase $\phi$ space) 
to the shape of the leading peak and its wings (cf. Fig.\ref{fig2}). 
Such effect is not possible in the case of the outer-gap
model (cf. Fig.\ref{fig3}), nor it is possible in the polar cap model (see Wo\'zna et al. 2002); these two models
lead to
approximate `mirror'-symmetry 
in the double-peak lightcurves. 
In principle then, this property might also be used to 
discriminate between the two-pole caustic model and other models. 

An important testing-ground for any models will be polarization properties of the high-energy radiation.
For the time being, good-quality polarization information is available  only
for optical light from Crab (Smith et al.~1988).
An argument in favor of the caustic origin of the optical peaks of the Crab pulsar
is that the degree of polarization
drops to minimum values at the phases of both peaks
(cf.~Fig.~4c of Smith et al.~1988). Such
a drop is justified by virtue of the caustic nature of the peaks: it results from
a pile-up of polarized radiation with different polarization angles.

Smith et~al.~\cite{sm88} emphasize, that the behaviour of the polarization
as a function of rotation-phase
for Crab is strikingly similar for both peaks, i.e. the polarization 
behaviour at the phase of the leading peak repeats at the trailing peak.
The outer gap model is able to reproduce this feature, 
even though
each of the two peaks arises from a very different type of
caustic in this model: the leading peak is due to the caustic formed close to the
light cylinder at the {\it leading}
part of the emission region, whereas the trailing peak is due to the
caustic formed within the {\it trailing} part. 
Romani \& Yadigaroglu (1995) consider the ability to reproduce the double sweep in the polarization position 
angle to be one of major successes of the outer-gap model.
In the two-pole caustic model, the two peaks arise due to crossing
the same  type of caustic - the dominant caustic associated with the trailing part of the emission region (cf. section 2). 
We suspect, therefore, that such a double sweep should even more naturally
be produced by the two-pole caustic model. 
A comprehensive treatment of
the polarization properties of high-energy radiation in the two-pole caustic model will be the subject
in our future work.

We emphasize that  
the characteristic form of the high-energy lightcurves of pulsars
(the double-peak structure with bridge emission in high-energy, preceded by the peak in radio)
is an inherent property
of a rotating source with a magnetic dipole, 
with roughly uniform high-energy emissivity 
along the last open field lines.
Two recently proposed models
may provide physical grounds for the geometry of the two-pole caustic model:
the slot-gap model of Arons \& Scharlemann (1979), in the modern version
of Muslimov \& Harding (2003);
and the model of Hirotani et al. (2003) of an outer gap extended on either side of the 
null surface due to the currents.
When a realistic physical model for the extended gaps becomes available, 
calculations of spectral characteristics
within our model will be possible.

\acknowledgments
We acknowledge fruitful discussions with A.~Harding, K.~Hirotani 
and A.~Muslimov. 
We thank J.~Gil for bringing our attention to the problem of missing wings in 
the double-peak lightcurves in the outer-gap model
and R.~W.~Romani for pointing out some shortcomings
of the two-pole caustic model. 
Comments by the anonymous referee helped us to clarify the paper significantly.
This work was supported by the grant PBZ-KBN-054/P03/2001.
Part of the work was performed while JD held a National Research
Council Research Associateship Award at NASA/GSFC.

\clearpage


\begin{figure}
\epsscale{0.7}
\plotone{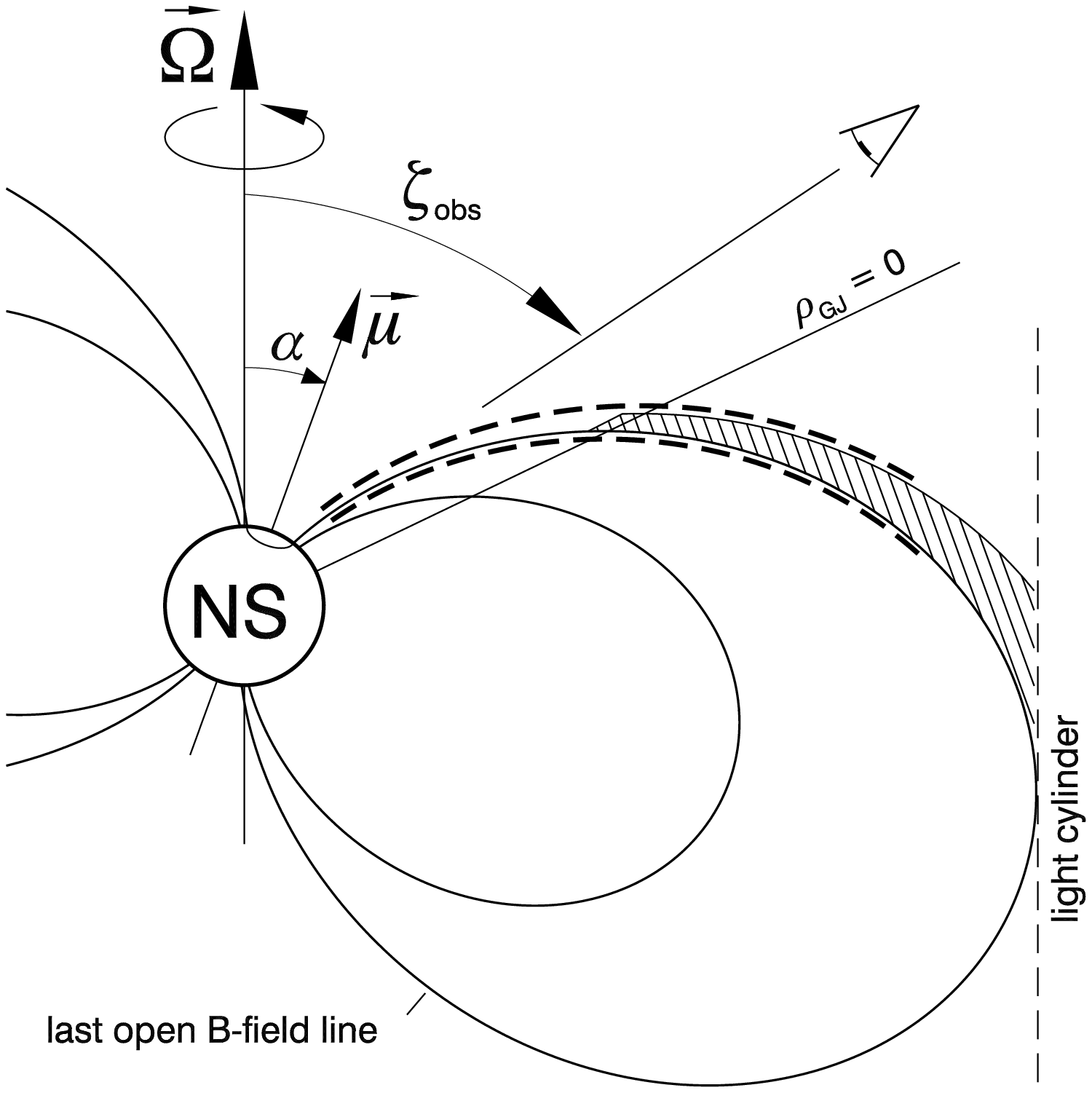}
\caption{Illustration of the two-pole caustic model. The radiating region
(within the dashed lines) is confined to the surface of the last closed 
field lines, and it
extends from the polar cap to the light cylinder. 
For comparison, the conventional
outer gap region is shown (shaded area), extending from the surface 
of the null space-charge 
($\rho_{\rm GJ} = 0$, where \hbox{$\rho_{\rm GJ} \approx 
-\vec \Omega \cdot \vec B(2\pi c)^{-1}$} is the Goldreich-Julian
charge density) to the light cylinder.
\label{fig1}}
\end{figure}

\clearpage 

\begin{figure}
\epsscale{0.6}
\plotone{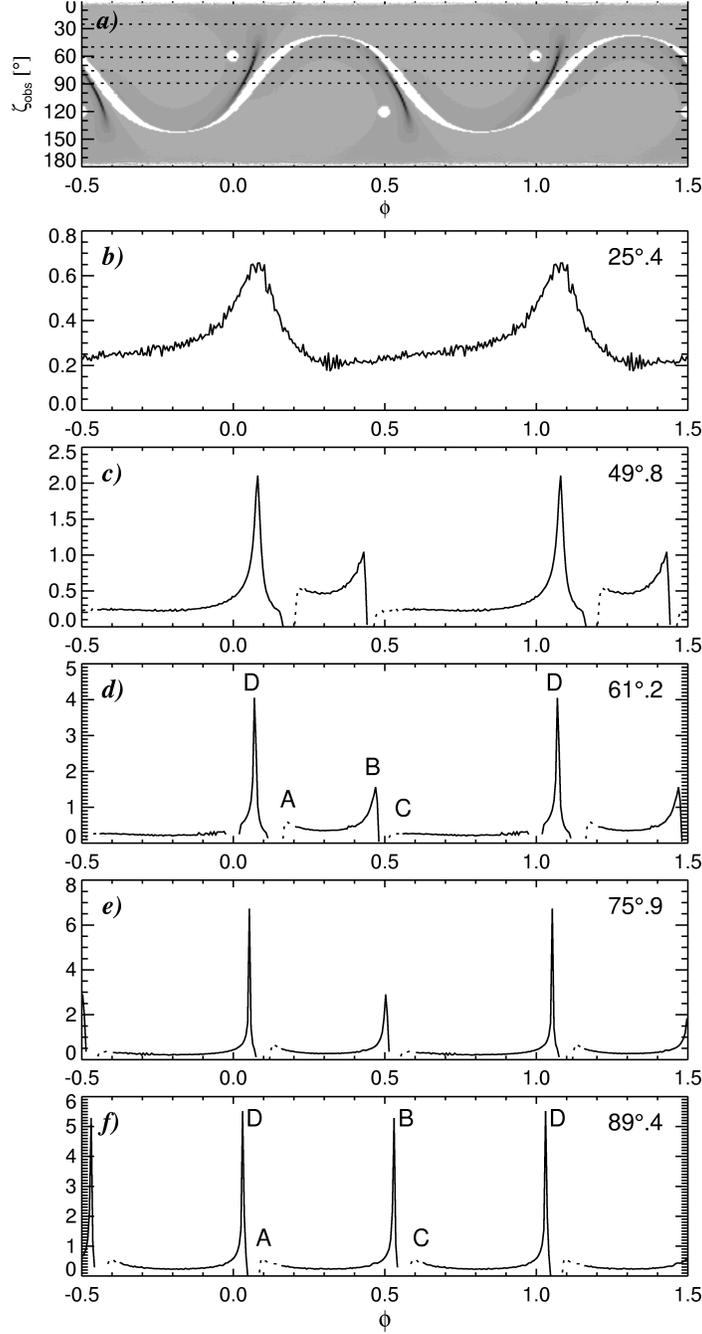}
\caption{Two-pole caustic model results:
Panel a): Photon mapping in the $(\zeta_{\rm obs}, \phi)$-plane for the inclination angle $\alpha = 60^\circ$.
Notice the location of the polar caps (the white spots; in this particular case
their size corresponds to the rotation period $P = 0.033$~s) 
as well as
two dominant caustics formed in the trailing parts of the magnetosphere (with respect to two magnetic poles).
Panels b) - f) show the lightcurves for five  viewing angles $\zeta_{\rm obs}$ 
given in the upper-right corners. Photon numbers are in arbitrary units. Within the narrow phase intervals at
A and C the lightcurves are drawn with dotted line in order to indicate lower accuracy there due to the proximity
of the light cylinder.}
\label{fig2}
\end{figure}

\clearpage 

\begin{figure}
\plotone{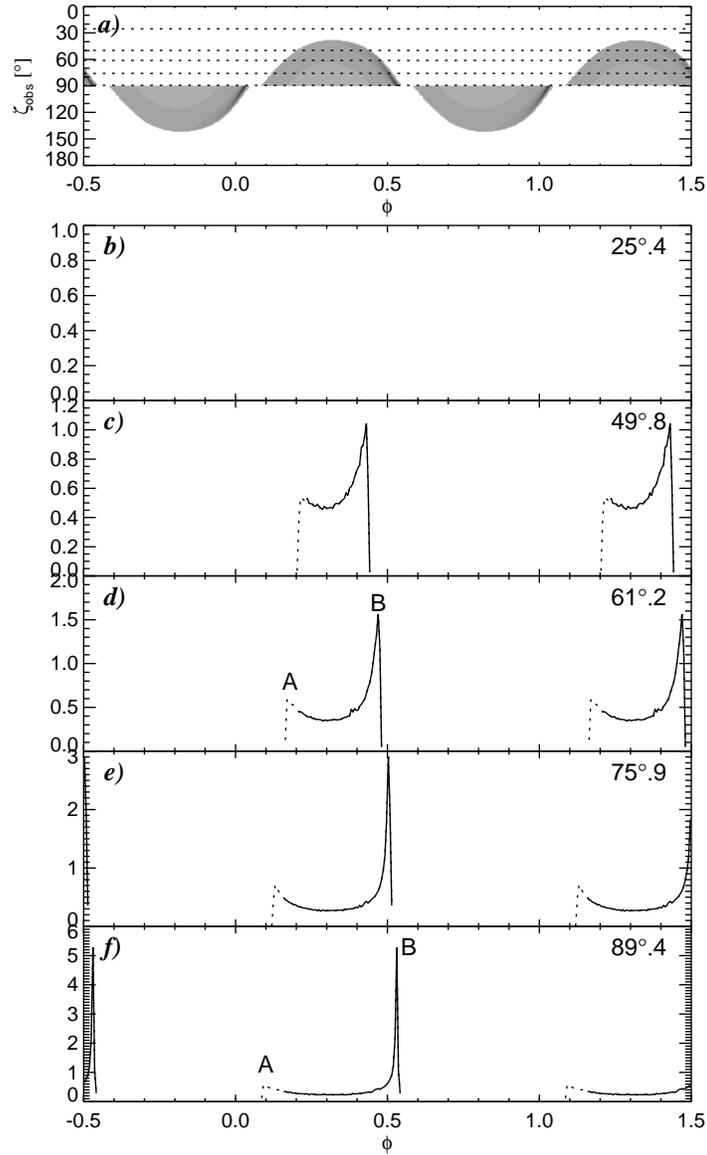}
\caption{The ``outer-gap" part of the model from Fig.\ref{fig2}: i.e. only those photons are shown which originate within the 
conventional outer gap region (shaded area in Fig.\ref{fig1}), all other photons have been removed. 
The ordinates in panels b)-f) were rescaled in order to enlarge the double-peak 
structure consisting now of A and B only.}
\label{fig3}
\end{figure}

\begin{figure}
\plotone{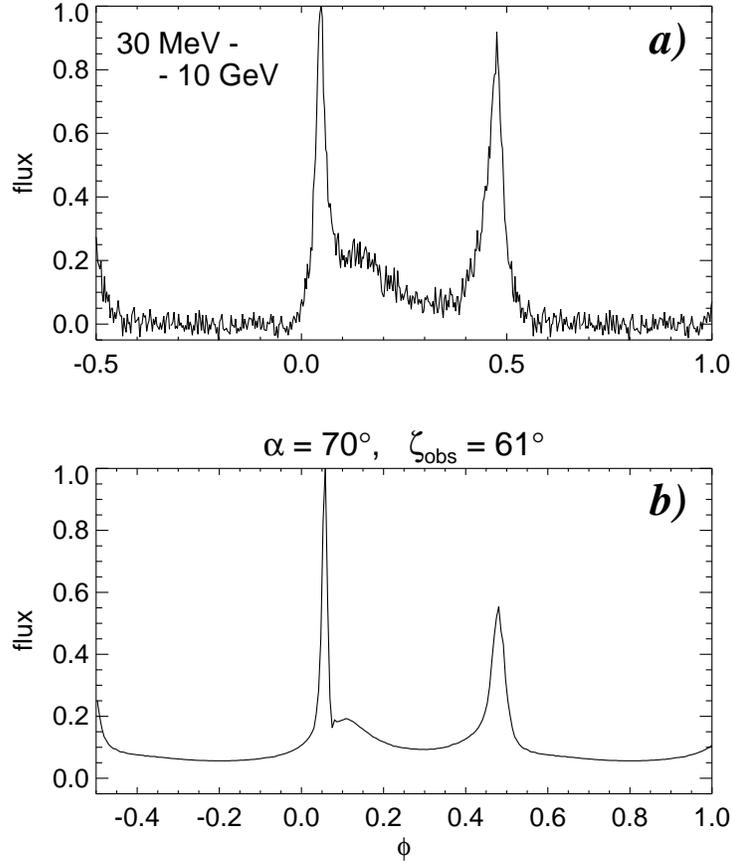}
\caption{
Comparison of the lightcurve predicted by the two-pole caustic model (panel b)
with the gamma-ray lightcurve of the Vela pulsar obtained with EGRET (Kanbach 1999) (panel a).
The model lightcurve was calculated for $\alpha = 70^\circ$ and $\zeta_{\rm
obs}=61^\circ$. The flux on the vertical axis is in arbitrary units.
}
\label{fig4}
\end{figure}

\begin{figure}
\plotone{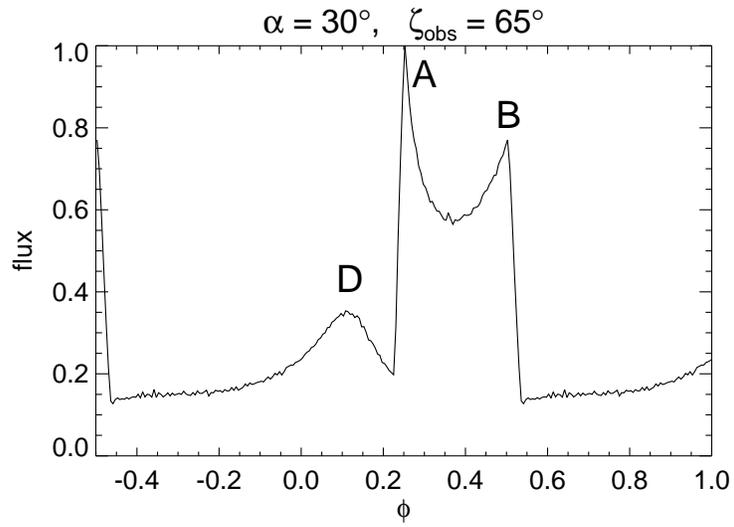}
\caption{
An example of a two-pole caustic model lightcurve 
dominated by the outer gap part of radiation
pattern (between A and B). The lightcurve was calculated for 
$\alpha = 30^\circ$ and $\zeta_{\rm obs} = 65^\circ$. 
}
\label{fig5}
\end{figure}

\clearpage

\begin{table}
\tabletypesize{\scriptsize}
\begin{center}

\caption{Comparison of models}
\vskip2mm
\begin{tabular}{|l | c c c c |c|}
\tableline

Characteristics of the model & $\begin{array}{c} \rm polar\\         
                 \rm cap  \end{array}$ & $\begin{array}{c} \rm Morini\\
                 \rm 1983  \end{array}$ & 
$\begin{array}{c} \rm Smith\ et\ al.\\
                 \rm 1988  \end{array}$

& 
            $\begin{array}{c} \rm outer\\             
                 \rm gap  \end{array}$ & 
                 $\begin{array}{c} \rm two-pole\\             
                 \rm caustic  \end{array}$\\

\tableline
caustic origin of the peaks & $-$ & $\pm$ ${\rm 2nd
\atopwithdelims ( ) \rm peak}$ & $-$ & + & + \\
\tableline
\hskip-1.7mm$\begin{array}{l} \rm each\ peak\\ 
                 \rm in\ the\ double-peak\ lightcurve\\
                 \rm is\ associated\ with\\
                \rm a\ different\ magnetic\ pole\end{array}$ 
                 & $-$ & $-$ & + & $-$ & + \\
\tableline
\hskip -1.7mm$\begin{array}{l} \rm photons\ emitted\ along\ the\ entire\\
\rm length\ of\ the\ magnetic\ field\ lines \end{array}$ 
    & + & + & $-$ & $-$ & + \\
\tableline
(the acceleration region is extended) & $-$ & $-$ & $-$ & + & + \\
\tableline
\end{tabular}

\end{center}
\end{table}

\end{document}